%% file: ML_PRF.tex
\documentclass[aip,pof,reprint,onecolumn,nofootinbib]{revtex4-1}

\usepackage{amsmath}
\usepackage{amssymb}
\usepackage{graphics}
\usepackage{graphicx}
\usepackage{psfrag}
\usepackage{subfigure}
\usepackage{color}
\usepackage{url}
\usepackage{enumerate}
\usepackage{rotating}
\usepackage{fancyhdr}
\usepackage[ruled, vlined]{algorithm2e}
\usepackage[symbol]{footmisc}

\usepackage[normalem]{ulem}
\usepackage{color}

\definecolor{gray}{rgb}{0.6,0.6,0.6}

\renewcommand{\thefootnote}{\fnsymbol{footnote}}

\makeatletter
\def\@dotsep{4.5}
\makeatother

\input{symbollines.tex}

\begin{document}

\title[Predictions of turbulent shear flows using deep neural networks]{Predictions of turbulent shear flows using deep neural networks}

\author{P. A. Srinivasan}
\affiliation{\mbox{Linn\'e FLOW Centre, KTH Mechanics, Stockholm, Sweden}}
\affiliation{\mbox{School of Electrical Engineering and Computer Science, KTH, Stockholm, Sweden}}
\affiliation{\mbox{Swedish e-Science Research Centre (SeRC), Stockholm, Sweden}}
\author{\textcolor{black}{L. Guastoni}}
\affiliation{\mbox{Linn\'e FLOW Centre, KTH Mechanics, Stockholm, Sweden}}
\affiliation{\mbox{Swedish e-Science Research Centre (SeRC), Stockholm, Sweden}}
\author{H. Azizpour}
\affiliation{\mbox{School of Electrical Engineering and Computer Science, KTH, Stockholm, Sweden}}
\affiliation{\mbox{Swedish e-Science Research Centre (SeRC), Stockholm, Sweden}}
\author{P. Schlatter}
\affiliation{\mbox{Linn\'e FLOW Centre, KTH Mechanics, Stockholm, Sweden}}
\affiliation{\mbox{Swedish e-Science Research Centre (SeRC), Stockholm, Sweden}}
\author{R. Vinuesa}
\email{rvinuesa@mech.kth.se}
\affiliation{\mbox{Linn\'e FLOW Centre, KTH Mechanics, Stockholm, Sweden}}
\affiliation{\mbox{Swedish e-Science Research Centre (SeRC), Stockholm, Sweden}}

\begin{abstract}
In the present work we assess the capabilities of neural networks to predict temporally evolving turbulent flows. In particular, we use the nine-equation shear flow model by Moehlis {\it et al.} [New J. Phys. {\bf 6}, 56 (2004)] to generate training data for two types of neural networks: the multilayer perceptron (MLP) and the long short-term memory (LSTM) network. We tested a number of neural network architectures by varying the number of layers, number of units per layer, dimension of the input, weight initialization and activation functions in order to obtain the best configurations for flow prediction. Due to its ability to exploit the sequential nature of the data, the LSTM network outperformed the MLP. The LSTM led to excellent predictions of turbulence statistics (with relative errors of $0.45\%$ and $2.49\%$ in mean and fluctuating quantities, respectively) and of the dynamical behavior of the system (characterized by Poincar\'e maps and Lyapunov exponents). \textcolor{black}{ This is an exploratory study where we consider a low-order representation of near-wall turbulence. Based on the present results, the proposed machine-learning framework may underpin} future applications aimed at developing accurate and efficient data-driven subgrid-scale models for large-eddy simulations of more complex wall-bounded turbulent flows, including channels and developing boundary layers.
\end{abstract}

\maketitle

\textcolor{black}{\section{Introduction}}

Artificial neural networks are computational frameworks for learning specific tasks solely from examples, and are a popular method within the general area of machine learning. They were inspired by their biological counterparts, which are organized as hierarchical layers of neurons as discovered by Hubel and Wiesel.\cite{hubel_wiesel} 
Although in existence for several decades, the recent success of deep neural networks\footnote[1]{In this manuscript, artificial neural networks, deep neural networks and neural networks are used interchangeably.}(DNNs) can be attributed to the increased computational power (mainly through graphics processing units or GPU) and the generation of large-scale datasets which benefit the training of the overparametrized DNNs. As discussed in the recent review by Duraisamy {\it et al.},\cite{duraisamy_et_al} there have been several studies aimed at using machine learning to model turbulent flows, especially in the context of Reynolds-averaged Navier--Stokes (RANS) simulations.\cite{ling_et_al,wu_et_al} Other applications to modelling the near-wall region of turbulent flows were reported by Milano and Koumoutsakos,\cite{milano_koumoutsakos} while Lapeyre {\it et al.}\cite{lapeyre_et_al} and Beck {\it et al.}\cite{beck_et_al} have documented the possibility of using machine learning in designing subgrid-scale (SGS) models for large-eddy simulation (LES). Machine learning has also been applied to flow control,\cite{kim_nn,gautier_et_al} development of low-dimensional models,\cite{shimizu_kawahara} generation of inflow conditions\cite{fukami_et_al} or structure identification in two-dimensional decaying turbulence,\cite{jimenez_ml} and as discussed by Kutz\cite{kutz} DNNs will progressively be more widely used in fluid mechanics in the coming years.

In this study we assess the feasibility of using DNNs to predict the temporal dynamics of simple turbulent flows. If such predictions are satisfactory, then DNNs could be employed for instance as SGS models in LES of more complex cases. As a simplified yet fairly complete model for wall-bounded turbulence, we consider the nine-equation model of a shear flow between infinite parallel free-slip walls subjected to a sinusoidal body force developed by Moehlis {\it et al.}\cite{moehlis_et_al} In this model, nine Fourier modes $\mathbf{u}_{j}(\mathbf{x})$ represent the mean profile, streamwise vortices, streaks and their instabilities, as well as the coupling among them. The instantaneous velocity fields in spatial coordinates $\mathbf{x}$ and time $t$ are then given by superposition of these nine modes as: 
\begin{equation} 
\textcolor{black}{\mathbf{\tilde{u}}(\mathbf{x},t) = \sum_{j=1}^9 a_{j}(t) \mathbf{u}_j(\mathbf{x}).}
\end{equation}
Through Galerkin projection, a system of nine ordinary differential equations (ODEs) for the nine mode amplitudes $a_{j}(t)$ is obtained. In this work, we considered a \textcolor{black}{ model} Reynolds number based on channel full height $2h$ and laminar velocity $U_{0}$ at a distance of $h/2$ from the top wall equal to $Re=400$, and we used $U_{0}$ and $h$ as velocity and length scales, \textcolor{black}{respectively}. \textcolor{black}{ The Fourier base modes $\mathbf{u}_{j}(\mathbf{x})$ are independent of the model Reynolds number, which defines the viscosity and therefore the dissipation of the problem.} The ODE model was then employed to produce time series of the nine amplitudes, to be used as training and validation datasets. Note that, as discussed by Moehlis {\it et al.},\cite{moehlis_et_al} the lifetime of the turbulent state is highly sensitive to the initial conditions and domain size. We produced over 10,000 different turbulent time series with a time span of 4,000 time units (where the time unit is defined in terms of $U_{0}$ and $h$), and considered a constant time step between samples of 1 time unit. The following initial conditions were employed for the Fourier modes:  $(a_{1}^{0},a_{2}^{0},a_{3}^{0},a_{5}^{0},a_{6}^{0},a_{7}^{0},a_{8}^{0},a_{9}^{0}) = (1,0.07066,-0.07076,0,0,0,0,0)$, as in the work by Kim\cite{kim_thesis} (here the superscript 0 denotes a value at $t=0$). The value of $a_{4}^{0}$ was randomly perturbed around 0, and if the generated series reached a fixed point or a periodic orbit then it was discarded. \textcolor{black}{According to Moehlis {\it et al.}\cite{moehlis_et_al}, higher \textcolor{black}{ model} Reynolds numbers lead to longer lifetimes of the perturbation with respect to the laminar profile \textcolor{black}{ (a result consistent with the lower dissipation)}.  However, in this study we consider only time series that are turbulent over the whole time span, and for this reason the present analysis would not depend on the particular \textcolor{black}{ model} Reynolds number under consideration.} We considered a domain with $L_{x}=4 \pi$, $L_{y}=2$ and $L_{z}=2 \pi$, where $x$, $y$ and $z$ are the streamwise, wall-normal and spanwise coordinates. In this study we assess flow predictions obtained with \textcolor{black}{two different types of neural network}, and use the machine learning software framework developed by Google Research called TensorFlow.\cite{tensor_flow} 

\textcolor{black}{This manuscript is organized as follows: in $\S$\ref{sec:mlp} we describe the characteristics of the multilayer perceptron (MLP), and report the results obtained with this type of network; in $\S$\ref{sec:lstm} we provide a description of the long short-term (LSTM) networks, and discuss their performance both in terms of turbulence statistics and dynamical behavior of the system; finally, in $\S$\ref{sec:conclusions} we summarize the main conclusions of the study.}

\textcolor{black}{\section{Predictions based on multilayer perceptrons (MLPs)} \label{sec:mlp}} 

\begin{figure}
\centering
\includegraphics[width=0.58 \textwidth]{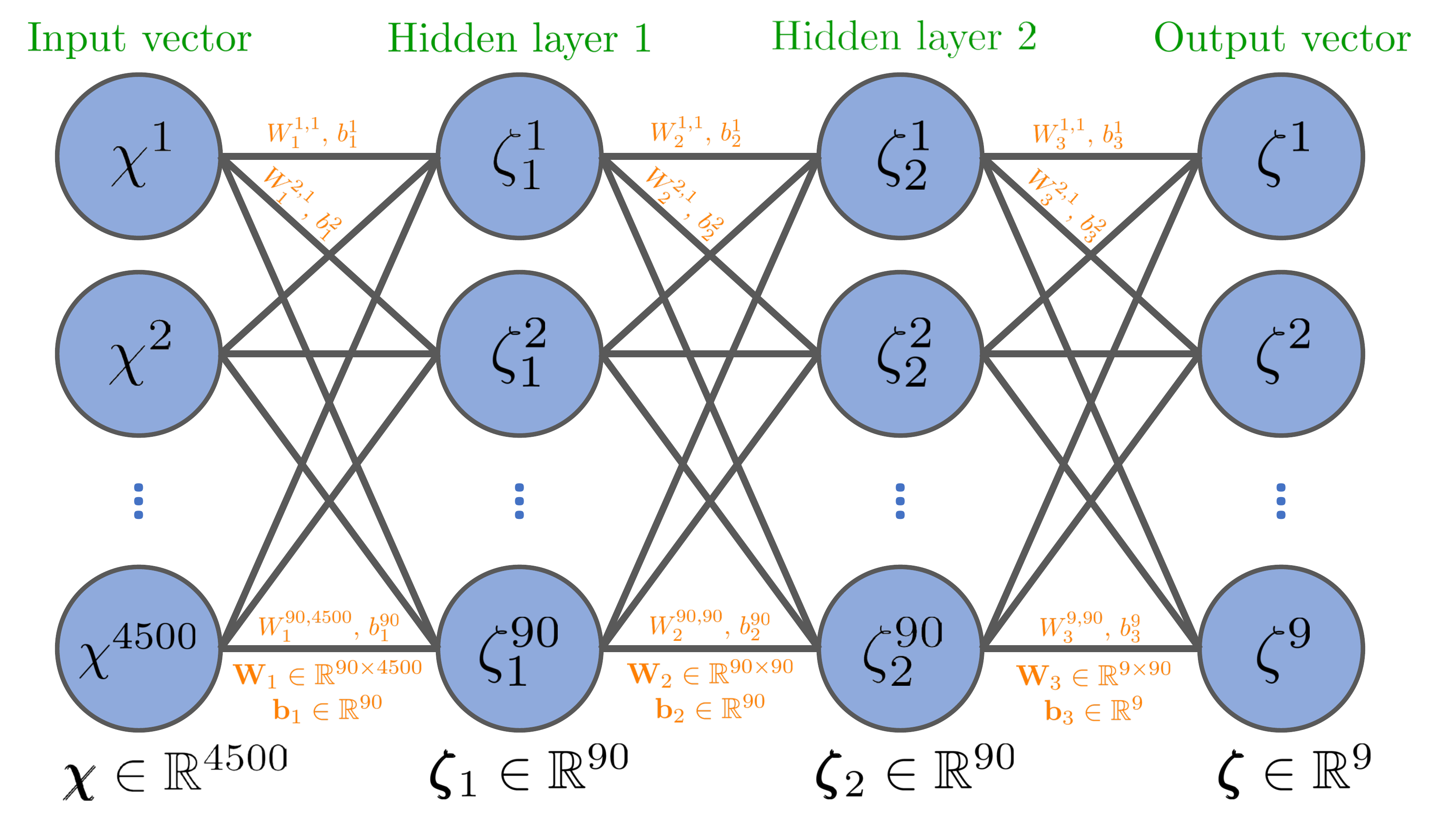}
\includegraphics[width=0.41 \textwidth,trim={3cm 1cm 3cm 0},clip=true]{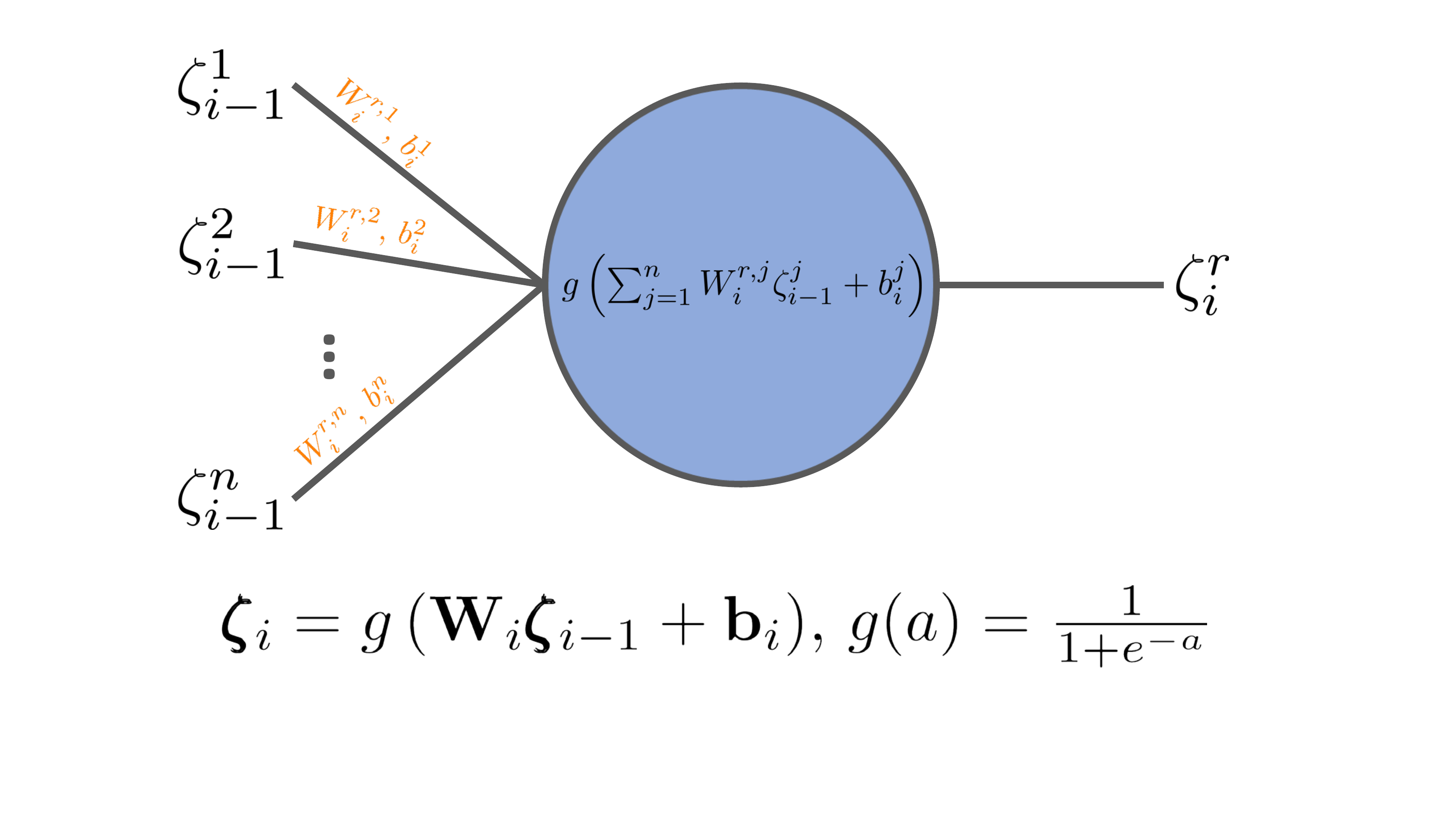}
\caption{(Left) Schematic representation of a multilayer perceptron (MLP) with two hidden layers, {\it i.e.} $l=2$, where each circle in the diagram depicts a single neuron. The MLP takes as input a $d$-dimensional vector $\pmb{\chi}$, then produces hidden $n$-dimensional vectors $\pmb{\zeta}_{i}$ for each of the two hidden layers, and finally generates an output vector $\pmb{\zeta}$ with $m$ dimensions. Note that the dimension of the input depends on the number of previous temporal values used for the prediction $p$. In this example $p=500$, $d=9p=4500$, $m=9$ and $n=90$, which implies that 90 neurons are used for each of the hidden layers. The transformation at a certain layer $i$ is performed with the model parameters given by the weight matrix $\mathbf{W}_{i}$. The $\mathbf{W}_{i}$ values are determined through learning using a gradient-following optimization technique on training data. (Right) Detail of a neuron, which takes a vector as input and generates a scalar output by performing a vector dot product and a nonlinear operation given by the activation function $g$.}
\label{fig_MLP}
\end{figure}

First, we consider multilayer perceptrons (MLPs),\cite{rumelhart1985learning} which are the most basic type of artificial neural network. As shown schematically in Figure~\ref{fig_MLP}, they consist of two or more layers of nodes (also called neurons), with each node connected to all nodes in the preceding and succeeding layers. Neural network training is based on the back-propagation algorithm,\cite{rumelhart_et_al} which operates layer by layer and updates the parameters in each layer's nodes to improve the predictions on the training data \textcolor{black}{using gradient-following methods.} If $X$ and $\Psi$ are input and output spaces, and each pair of vectors $(\pmb{\chi},\pmb{\psi})$ is a training example or sample, the objective of the neural network is to find the mapping ${f: X \rightarrow \Psi}$ using the training set such that a loss function $L(f(\pmb{\chi}); \pmb{\psi})$ is minimized. For an MLP, the mapping $f(\pmb{\chi})$ can be decomposed into a sequence of simple transfer functions \textit{i.e.} linear matrix transform followed by an element-wise nonlinear function called activation function (\textit{e.g.}\ sigmoidal). Figure~\ref{fig_MLP} summarizes the basic structure of an MLP, where $l$ is the number of hidden layers, $n$ the number of neurons per hidden layer (which in our case will be constant for all the hidden layers) and $p$ is the number of previous values used to predict the next one. The evaluation of $\pmb{\zeta} = f(\pmb{\chi})$ is done using Algorithm~\ref{algo_mlp}, which involves the following parameters for each layer $i$: $g_{i}$ are the activation functions (which filter the contributing units in each layer), whereas the weight matrices $\mathbf{W}_i$ and biases $\mathbf{b}_i$ perform a linear transformation from one hidden layer space to the next. The goal of training an MLP is to determine $\mathbf{W}_i$ and $\mathbf{b}_i$ using training data, for the given activation functions. The back-propagation algorithm used to update the weights and biases involves the use of gradient descent, for which a so-called learning rate (step size) needs to be defined. The learning rate can be manually set to a constant value or allowed to decay over time, since an excessively large learning rate may prevent convergence. For all the training runs in this work, a variant of the stochastic gradient descent algorithm called adaptive moment estimation (Adam)\cite{kingma_ba} is used. Adam has an adaptive learning rate method which is commonly used to train deep networks.

\begin{algorithm}
\DontPrintSemicolon
\KwIn{$\pmb{\chi}$}
\KwOut{$\pmb{\zeta} = f(\pmb{\chi})$}
 set $\pmb{\zeta}_{0}  \leftarrow \pmb{\chi}$\;
 \For{$i\leftarrow 1$ \KwTo $l+1$}{
  \ \ $\mathbf{h}_i \leftarrow \mathbf{W}_{i} \pmb{\zeta}_{i-1}  + \mathbf{b}_i$\;
  \ \ $\pmb{\zeta}_{i}  \leftarrow g_i(\mathbf{h}_i)$\;
 }
 return $\pmb{\zeta} \leftarrow \pmb{\zeta}_{l+1}$\;
 \caption{Compute MLP output.}
 \label{algo_mlp}
\end{algorithm}

\begin{figure}[t!]
\centering
\vspace{0pt}
\includegraphics[width=0.75 \textwidth]{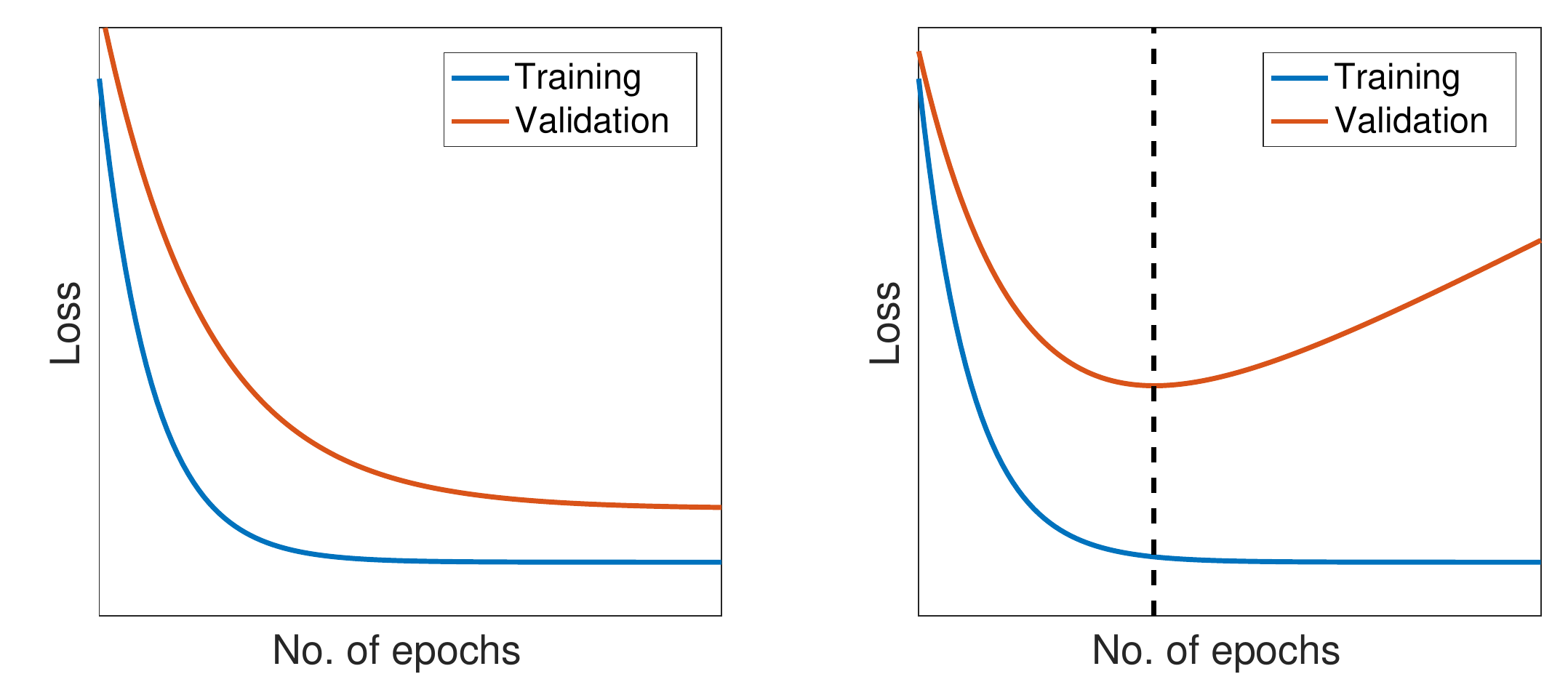}
\caption{\textcolor{black}{Schematic representation of the training and validation losses as a function of the number of epochs. (Left) Adequate fitting, where the validation loss does not increase with the number of epochs, and (right) shows overfitting beyond the dashed vertical line (which indicates the point for early stopping).}}
\label{fig_over}
\end{figure}

The updates for the weights and biases in the algorithm depend on the choice of the loss function, and a standard choice for regression problems is the mean-squared-error. Since training a neural network is essentially fitting a mathematical function to data, it inherently has a risk of overfitting. An overfitted network is one that has achieved very low losses and performs well on the training set, but performs poorly on new data. To overcome this problem, we use a regularization technique which penalizes too large parameters based on their $L_2$ norm. It is also common to set aside a small portion of the training set as a validation set which will not be used during training. While training, one complete pass through the training set is called an epoch. At the end of each epoch, the model is evaluated using the validation set and a validation loss is calculated using the loss function. A common method to avoid overfitting is called early stopping, whereby the training is stopped when the validation loss starts increasing. \textcolor{black}{This method is illustrated in Figure~\ref{fig_over}.} Other relevant decisions to be made when training deep learning models are the strategy for weight initialization and the choice of activation functions. Since the MLP architecture has a large number of weights, the training algorithm seeks to find a local minimum in a high-dimensional space, and the rate of convergence of this search, among other factors, depends on the initial values of these weights. Although randomly selecting initial weights based on a uniform distribution is a natural choice, there are a few other options that may perform better. For instance, since training a neural network involves a forward pass coupled with back-propagation, it is generally better to prevent successive enlargement or shrinkage of layer outputs. Hence, initializations based on a suitable normal distribution perform better than those based on uniform distributions. A truncated normal distribution, which bounds the random variable both above and below, is a commonly used distribution for most learning tasks. Other variations based on the normal distribution, showing better convergence properties than the truncated normal, were proposed by Glorot and Bengio\cite{glorot_bengio} and He {\it et al.}\cite{he_et_al} Regarding the activation function, the most widely used is the hyperbolic tangent, although other alternatives (see for instance the work by LeCun {\it et al.}\cite{lecun_et_al}) such as the rectified linear unit (ReLU),\cite{nair2010rectified} the Leaky ReLU\cite{maas2013rectifier} and the exponential linear unit (eLU)\cite{clevert2015fast} are also reported in the literature. 

For the predictions of the near-wall turbulence model using MLPs, we consider the mean-squared-error as loss function, and define the validation loss as:
\begin{equation}
\textcolor{black}{L(f(\pmb{\chi});\pmb{\psi})=\frac{1}{2m} \sum_{j=1}^{m} \left | \psi^{j}-f(\chi^{j}) \right |^{2},}
\end{equation}
\textcolor{black}{where $\pmb{\psi}$ and $f(\pmb{\chi})$ are the $m$-dimensional vectors containing the true and predicted values, respectively.} We use $20\%$ of the training data as a validation set, and consider early stopping to avoid overfitting. The various initialization strategies were compared in the study by Srinivasan,\cite{premsrinivasan} and the one by Glorot and Bengio\cite{glorot_bengio} was observed to outperform all the others. Therefore, that specific initialization is employed in this study. Srinivasan\cite{premsrinivasan} also compared the various activation functions, and the hyperbolic tangent provided the best agreement with reference data, therefore it was employed for the MLP predictions reported below. The next step to design an MLP architecture is to decide the number of hidden layers $l$, the number of units per hidden layer $n$  and the dimensionality of the input $d$, which is the number of predicted quantities (9 in our case) multiplied by the number of previous values from each amplitude $a_{j}$ used to predict the next one, denoted by $p$. \textcolor{black}{We will start by assessing the value of $p$ required for the predictions, and to do it we recall that, as indicated in Figure~\ref{fig_over}, a properly trained neural network must be able to first reproduce the training data (which has been seen by the network during training), before attempting to reproduce the validation data (not seen during training). We train the MLP with a single time series of 4,000 time units for several epochs, and then predict the same dataset. For this test, we maintain a constant number of units per hidden layer of $n=90$, and vary the values of $p$ and $l$. We define the relative error in the prediction of amplitude $a_{1}$ (over $N_{s}$ samples) as:}
\begin{equation}\label{epsilon_eq}
\textcolor{black}{\varepsilon_{1}=\frac{1}{(N_{s}-p) a_{1,{\rm lam}} } \sum_{j=p+1}^{N_{s}} \left | a_{1,{\rm tra}}^{j}-a_{1,{\rm pred}}^{j} \right |,}   
\end{equation} 
\textcolor{black}{where `tra' and `pred' denote data from the training dataset ({\it i.e.}, seen by the network during training) and data predicted by the NN, respectively, and $a_{1,{\rm lam}}=1$ is the value for laminar flow. The values of $\varepsilon_{1}$ for the various cases are summarized in Table~\ref{table_epsilon}, which shows that, for the data under consideration in this study, it is necessary to consider a value of $p=500$ in order to obtain acceptable training losses, {\it i.e} below $10\%$ when at least 2 hidden layers are considered.}
\begin{table}
\caption{\textcolor{black}{Relative errors obtained when predicting a training dataset $\varepsilon_{1}[\%]$.}} 
\begin{center}
\begin{tabular}{c c c c c c}
\hline \hline
$p$ & $l=1$ & $l=2$ & $l=3$ & $l=4$ & $l=5$   \\[3pt]
\hline
10 & 23.34 & 33.66 & 25.37 & 26.01 & 23.81 \\
50 & 28.28 & 20.59 & 17.66 & 16.77 & 22.95 \\
100 & 18.97& 13.58 & 12.76 & 17.17 & 14.42 \\
500 & 65.21&  3.32 & 6.80  & 1.43  & 3.09 \\ 
 \hline \hline
\end{tabular}
\label{table_epsilon}
\end{center}
\end{table}

\textcolor{black}{Based on the results discussed above, we will consider in all the MLP cases a value of $p=500$, which leads to a reasonably large input dimension of $d=4,500$, while the output dimension is $m=9$. In order to select the rest of parameters, five different MLP architectures with various values of $n$ and $l$ were evaluated, as summarized in Table~\ref{table_mlp}.} After using 1,000 datasets (each of them consisting of a time series spanning 4,000 time units) to train the MLP, the predicted mode amplitudes were employed to reconstruct the velocity fields. These fields were used to calculate the mean velocity profile $\overline{u}(y)$ and the streamwise velocity fluctuations $\overline{u^{2}}(y)$, based on averaging over the two periodic directions $x$ and $z$, in time over the $4,000$ time units of a time series, and then performing an ensemble average over 500 different time series, which were sufficient to ensure statistical convergence.\cite{premsrinivasan} These profiles were compared with reference statistics obtained from the Moehlis {\it et al.}\cite{moehlis_et_al} model, also based on 500 time series which had not been seen by the MLP during training. We define the relative error between the model and the MLP prediction (denoted by the subindices `mod' and `pred', respectively) for the mean flow as:
\begin{equation}
\textcolor{black}{E_{\overline{u}}=\frac{1}{2\  {\rm max}(\overline{u}_{{\rm mod}})} \int_{-1}^{1} \left | \overline{u}_{{\rm mod}}-\overline{u}_{{\rm pred}} \right |  {\rm d}y,}
\end{equation}
where the normalization with the maximum of $\overline{u}$ is introduced to avoid spurious error estimates close to the centerline where the velocity is 0. This error is defined analogously for $\overline{u^{2}}$, and both are reported in Table~\ref{table_mlp} for each of the architectures, together with the validation loss defined above. These results show that in general MLPs with higher numbers of hidden layers $l$ lead to better predictions of the amplitudes and also of the turbulence statistics, whereas increasing $n$ does not always lead to better predictions. The best MLP architecture among the ones under study was MLP4, with 5 hidden layers and a total of 90 neurons per layer. Although the agreement with the reference mean velocity profile obtained with MLP4 is acceptable (with a relative error of $3.21\%$), the streamwise velocity fluctuation profile exhibits a much larger deviation with $18.61\%$ relative error. Given the relatively large size of the network, and particularly the high input dimension, we explored alternative DNNs which are able to better exploit the sequential nature of the data as discussed below.
\begin{table}
\caption{Summary of MLP architectures and their performance using 1,000 training datasets.} 
\begin{center}
\begin{tabular}{c c c c c c}
\hline \hline
Architecture & $l$ & $n$ & $E_{\overline{u}}$$[\%]$ & $E_{\overline{u^{2}}}$$[\%]$ & Validation Loss  \\[3pt]
\hline
MLP1 & 4 & 45  & 1.84 & 24.91 & $3.96 \times 10^{-5}$ \\
MLP2 & 3 & 90  & 10.96 & 36.16 &  $4.38 \times 10^{-5}$ \\
MLP3 & 4 & 90  & 7.00 & 29.04 & $3.90 \times 10^{-5}$ \\
MLP4 & 5 & 90  & 3.21 & 18.61 & $3.84 \times 10^{-5}$ \\
MLP5 & 4 & 180 & 5.87 & 27.85 & $3.99 \times 10^{-5}$ \\
 \hline \hline
\end{tabular}
\label{table_mlp}
\end{center}
\end{table}

\textcolor{black}{\section{Predictions based on long short-term memory (LSTM) networks} \label{sec:lstm}} 
\textcolor{black}{Although MLPs are frequently used in practice, their major limitation is that they are designed for point prediction as opposed to time-series prediction, which might require a context-aware method. Nevertheless, MLPs provide a solid baseline in machine-learning applications and thereby help verifying the need for a more sophisticated network architecture or lack thereof.} An alternative to the MLP is the recurrent neural network (RNN), which is a type of neural network specifically suited to learn from sequential data such as time series. These networks have a much closer resemblance to biological neural networks, since they also learn the temporal dynamics of the input sequence. As opposed to MLPs, RNNs map a sequence of data points to another sequence using recurrent connections in time as outlined in Algorithm~\ref{algo_rnn}. A standard RNN is a neural network containing a single hidden layer with a feedback loop. The output of the hidden layer in the previous time instance is fed back into the hidden layer along with the current input. An RNN can be thought of as a very deep MLP with one layer for each instance in the input sequence where all layers share the parameters. This sequential feeding of the input data makes the temporal relationship apparent to the network. An RNN is parametrized by three weight matrices ($\mathbf{W}_{h\chi}$, $\mathbf{W}_{hh}$ and $\mathbf{W}_{\zeta h}$) and two biases ($\mathbf{b}_h$ and $\mathbf{b}_{\zeta}$). In Algorithm~\ref{algo_rnn}, $g_h$ and $g_{\zeta}$ are the hidden and output activation functions, respectively. The loss function of an RNN for a single training example is the sum over $p$ time steps of the squared error. On the other hand, the weights and biases are computed by back-propagation through time.\cite{werbos} RNNs are particularly difficult to train especially for sequences with long-range temporal dependencies. Since the back-propagation algorithm uses the chain rule to compute the weight corrections, small gradients in the later layers restrict the rate of learning of earlier layers. For long sequences, this gradient can gradually vanish, a fact that prevents the neural network from learning any further. This is called the problem of vanishing gradients. Long short-term memory (LSTM) networks developed by Hochreiter and Schmidhuber\cite{hochreiter_schmidhuber} use a gating mechanism to actively control the dynamics of the recurrent connections and thereby mitigate the vanishing gradient issue. Thanks to the same gating mechanism, LSTMs can also model longer temporal dependencies than standard RNNs. The LSTM will be the RNN employed in this study.
\begin{algorithm}[htbp]
\DontPrintSemicolon
\KwIn{Sequence $\pmb{\chi}_{1}, \pmb{\chi}_{2}, \dots \pmb{\chi}_{p}$}
\KwOut{Sequence $\pmb{\zeta}_{1}, \pmb{\zeta}_{2}, \dots \pmb{\zeta}_{p}$}
 set $\mathbf{h}_0 \leftarrow 0$\;
 \For{$t\leftarrow 1$ \KwTo $p$}{
  \ \ $\mathbf{h}_t \leftarrow g_h(\mathbf{W}_{h \chi} \pmb{\chi}_{t} + \mathbf{W}_{hh} \mathbf{h}_{t-1} + \mathbf{b}_{h})$\;
  \ \ $\pmb{\zeta}_{t} \leftarrow g_{\zeta}(\mathbf{W}_{\zeta h} \mathbf{h}_t + \mathbf{b}_{\zeta})$\;
 }
 \caption{Compute the output sequence of an RNN.}
 \label{algo_rnn}
\end{algorithm}

\begin{figure}[t!]
\centering
\vspace{0pt}
\includegraphics[width=0.96 \textwidth]{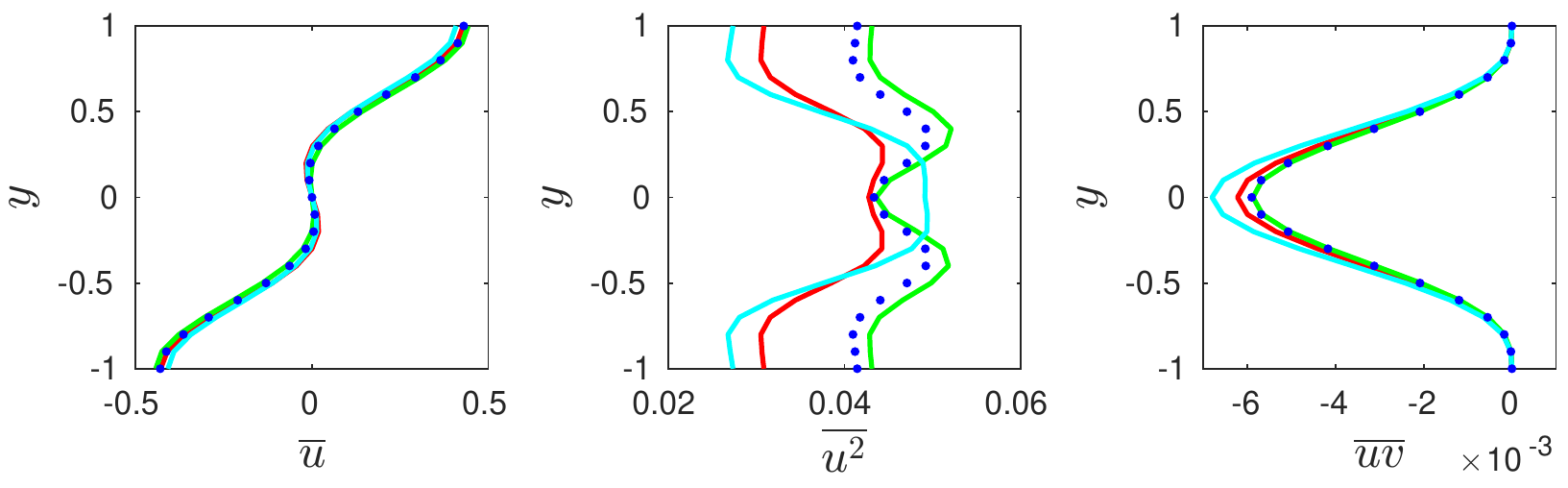}
\caption{\textcolor{black}{Turbulence  statistics  corresponding  to  (left)  streamwise  mean  profile,  (middle)  streamwise velocity  fluctuations  and  (right)  Reynolds  shear  stress. Blue dots are used for the reference nine-equation model, and colored lines denote predictions using the following architectures: (red) LSTM1, (green) LSTM2 and (cyan) LSTM3.  The networks were trained with 100 datasets.}}
\label{fig_lstm_comp}
\end{figure}

The impact of the considered LSTM architecture on the computed turbulence statistics is summarized in Table~\ref{table_lstm}, where the relative errors $E_{\overline{u}}$ and $E_{\overline{u^{2}}}$ are calculated as above based on ensembles of 500 time series spanning 4,000 time units each. Note that the LSTM1 and LSTM3 architectures have a single layer, while in LSTM2 we consider a second one, and in all the cases we have 90 units per layer. In this first assessment we only considered 100 datasets for training, and it can be observed how even using much less training data, \textcolor{black}{the LSTM1 and LSTM2 networks outperform all the MLPs discussed above, and the LSTM3 architecture yields results similar to those of the best MLP, {\it i.e.} MLP4. Furthermore,  the LSTM1 network} leads to relative errors lower than MLP4 using $p=10$, {\it i.e.}\ an input dimension 50 times lower than that in all the MLPs. \textcolor{black}{The mean velocity profile $\overline{u}$, together with the streamwise velocity fluctuations $\overline{u^{2}}$ and the Reynolds shear stress $\overline{uv}$ (which is associated to turbulent transport), are shown in Figure~\ref{fig_lstm_comp} for the three LSTM networks trained with 100 datasets. These are compared with the reference profiles obtained from the nine-equation model. This figure, together with the results in Table~\ref{table_lstm}, indicate that it would be possible to further improve the predictions from the LSTM1 network by considering a second layer as in the LSTM2 architecture. This significantly improves the prediction of the streamwise velocity fluctuation profile, reducing the relative error by about a factor of 2. Nevertheless, increasing the dimension of the input to $p=25$ appears to have a detrimental effect on the results, in particular when predicting the fluctuations. We observed that the deviations in the $\overline{u^{2}}$ predictions are induced by a subset of the predicted time series, which lead to the different behavior of the fluctuations observed in Figure~\ref{fig_lstm_comp}. Therefore the use of long input sequences for predictions using the LSTM network may lead to higher errors in the predictions, especially in higher-order moments (note that the prediction of the mean flow is in reasonably good agreement with the reference).} 

\textcolor{black}{The mean velocity profile and the Reynolds shear stress are connected, as it can be shown from a streamwise momentum balance. This was used by Vinuesa {\it et al.}\cite{convergence_meccanica} in turbulent channel flows (where such a balance reveals that the total shear stress is a linear function) to define a criterion for convergence of turbulence statistics. In the present case,\cite{moehlis_et_al} the total shear stress $1/Re\  {\rm d} \overline{u} / {\rm d}y - \overline{uv}$ is given by a cosine of amplitude $2 \sqrt{2} \pi /(4Re)$. The total shear stress computed from the nine-equation model, averaged over 500 datasets, has a deviation of around $2\%$ with respect to this value, a fact that supports the statistical convergence of our results. Interestingly, the predictions from the LSTM1 and LSTM2 networks (trained with 100 datasets) shown in Figure~\ref{fig_lstm_comp} exhibit $4.7\%$ and $2.8\%$ deviation with respect to the reference amplitude, which suggests that these neural networks reproduced the link between the viscous and turbulent stresses. On the other hand, the LSTM3 network shows a significantly larger deviation of $43.5\%$, despite its reasonably good prediction of the mean velocity profile. We conjecture that it could be possible to further improve the predictions of neural networks by explicitly including the physical constraints present in the flow when designing the network.}

\begin{table}
\caption{Summary of LSTM architectures and their performance using different numbers of training datasets.} 
\begin{center}
\begin{tabular}{c c c c c c c c}
\hline \hline
Architecture & $p$ & $n_{1}$ & $n_{2}$ & Number of datasets & $E_{\overline{u}}$$[\%]$ & $E_{\overline{u^{2}}}$$[\%]$ & Validation Loss  \\[3pt]
\hline
LSTM1 & 10 & 90  & -- & 100 & 2.36 & 14.73 & $2.0 \times 10^{-8}$ \\
LSTM1 & 10 & 90  & -- & 1,000 & 0.83 & 3.44  & $8.5 \times 10^{-9}$ \\
LSTM1 & 10 & 90  & -- & 10,000 & 0.45 & 2.49  & $5.2 \times 10^{-9}$ \\
LSTM2 & 10 & 90  & 90 &100 & 1.94 & 6.82 & $2.4 \times 10^{-8}$ \\
LSTM3 & 25 & 90  & -- & 100 & \textcolor{black}{3.53} & \textcolor{black}{18.28} & $\textcolor{black}{7.4} \times 10^{-8}$ \\
 \hline \hline
\end{tabular}
\label{table_lstm}
\end{center}
\end{table}

\begin{figure}[t!]
\centering
\vspace{0pt}
\includegraphics[width=0.95 \textwidth]{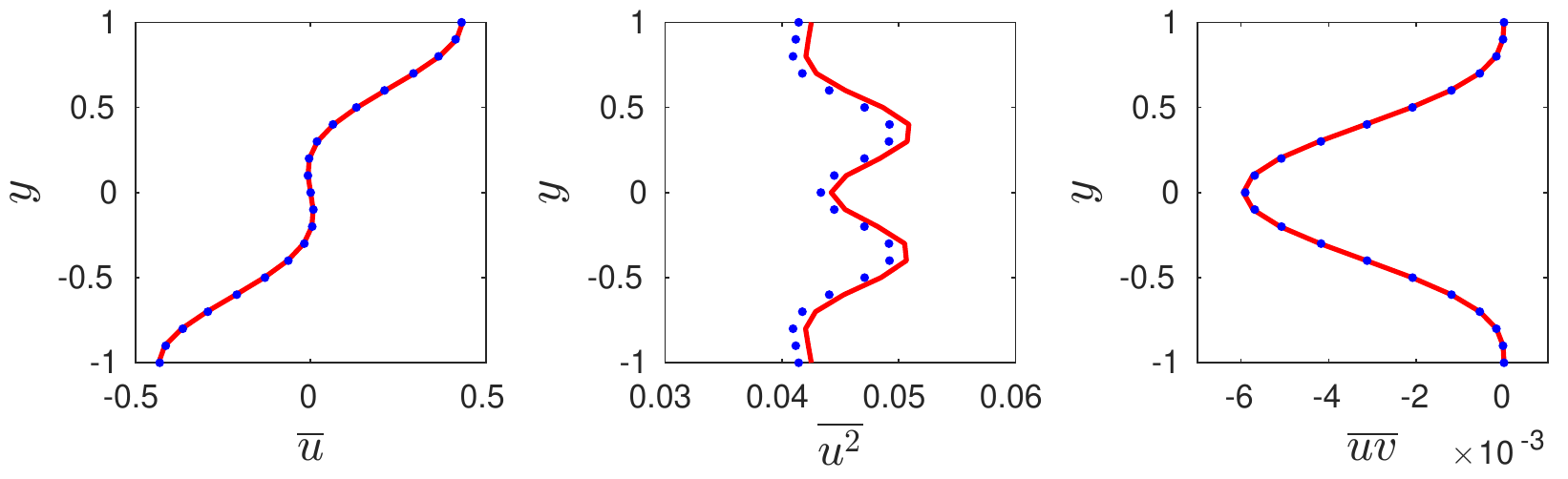}
\includegraphics[width=0.96 \textwidth]{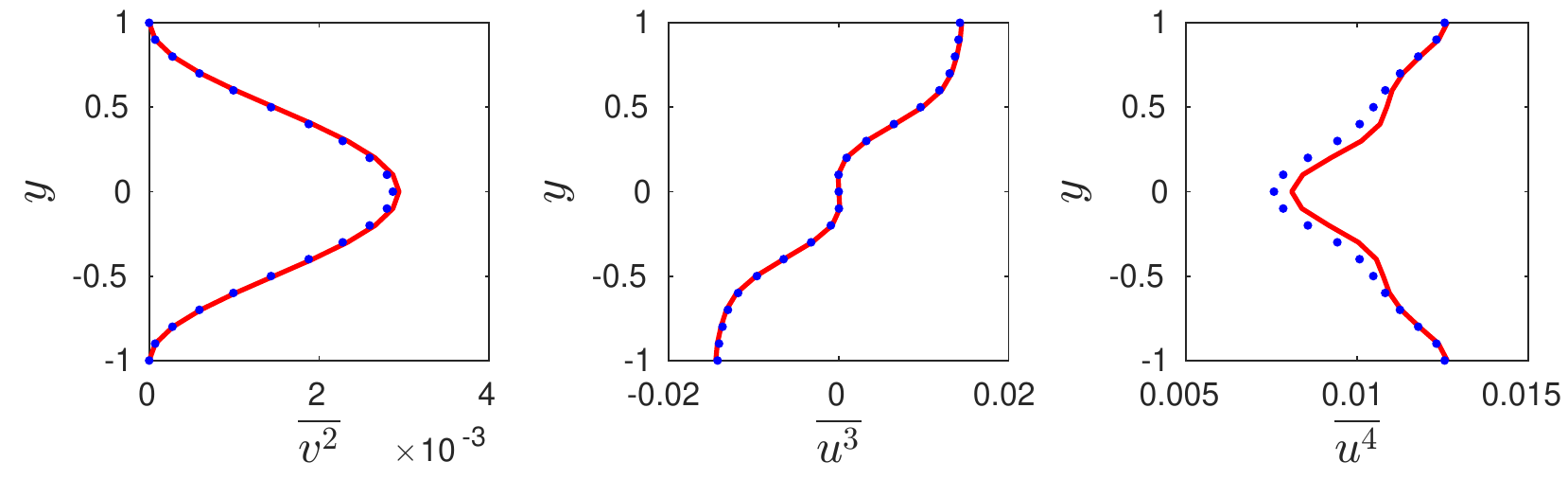}
\caption{\textcolor{black}{Turbulence statistics corresponding to (top-left) streamwise mean profile, (top-middle) streamwise velocity fluctuations and (top-right) Reynolds shear stress; (bottom-left) wall-normal velocity fluctuations, (bottom-middle) skewness and (bottom-right) flatness.} Blue dots are used for the reference nine-equation model and red lines for the predictions using the LSTM1 network trained with 10,000 datasets.}
\label{fig_stats}
\end{figure}

Furthermore, we also assessed the effect of increasing the number of datasets used for training, and calculated the deviations with respect to the reference turbulence statistics as summarized in Table~\ref{table_lstm}. \textcolor{black}{Although in order to limit the computational cost of the training we considered the smaller LSTM1 architecture with 1,000 and 10,000 training datasets, even lower error levels would be expected if the LSTM2 architecture was trained with larger amounts of data.} The data on this Table allows a direct comparison between the performance of the LSTM1 and MLP4 architectures, both trained with 1,000 datasets: while the LSTM network leads to relative errors of $0.83\%$ and $3.44\%$ in the mean and the fluctuations, the MLP \textcolor{black}{exhibits} significantly higher errors of $3.21\%$ and $18.61\%$, respectively. The LSTM1 results are in even better agreement with the reference statistics when 10,000 training datasets are employed for training, \textcolor{black}{with relative errors in the mean velocity and the streamwise fluctuations of $0.45\%$ and $2.49\%$, respectively. The excellent agreement between the turbulence statistics from the nine-equation model and the LSTM network can be observed in Figure~\ref{fig_stats}, where in addition to the mean streamwise velocity $\overline{u}$ and the streamwise velocity fluctuations $\overline{u^{2}}$, we show the Reynolds shear stress $\overline{uv}$, the wall-normal fluctuations $\overline{v^{2}}$, as well as the skewness $\overline{u^{3}}$ and the flatness $\overline{u^{4}}$. The agreement of all the statistics with the reference data is very good, and even higher-order moments like the skewness and the flatness exhibit low relative errors, {\it i.e.} $1.01\%$ and $2.57\%$, respectively. Also note that the amplitude of the cosine defining the total shear stress shows a deviation of $3.1\%$ with respect to the value expected from the streamwise momentum balance, a fact that suggests that this network also reproduces the connection between the viscous and turbulent stresses. It is interesting to note that the present results were obtained with a loss function based on minimizing the error in the instantaneous values of the mode coefficients, but other studies such as the one by King {\it et al.}\cite{king_et_al} propose to employ turbulence-specific loss functions; these have the potential of further improving the flow predictions. In any case, the present results} highlight the excellent predicting capabilities of the LSTM network, \textcolor{black}{which could potentially be further improved by considering a larger network.}

\begin{figure}[t!]
\centering
\includegraphics[width=0.49\textwidth]{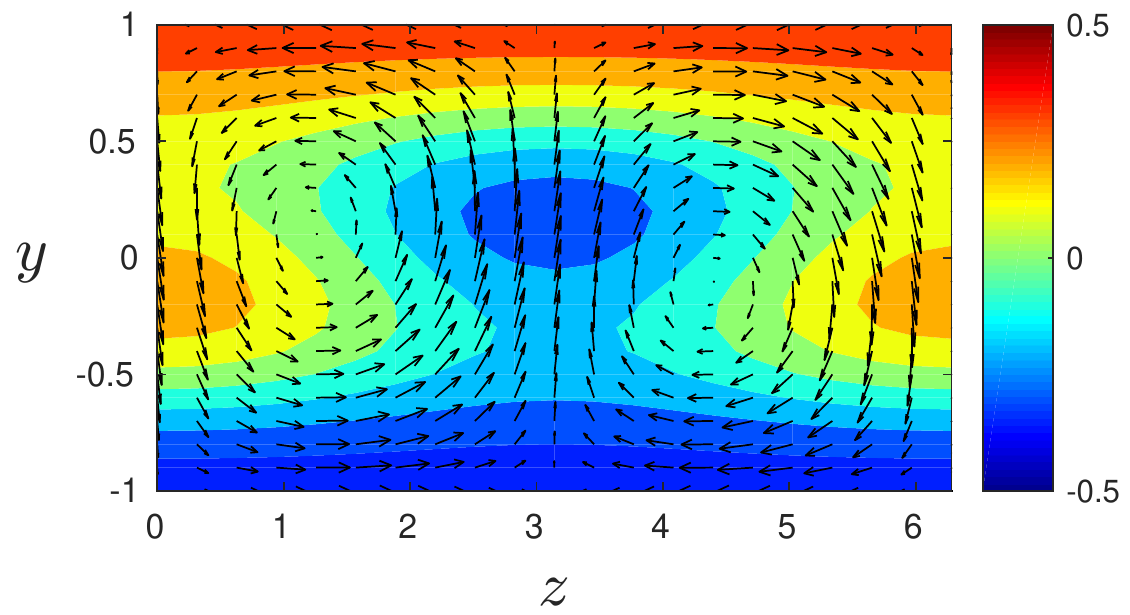}
\includegraphics[width=0.49 \textwidth]{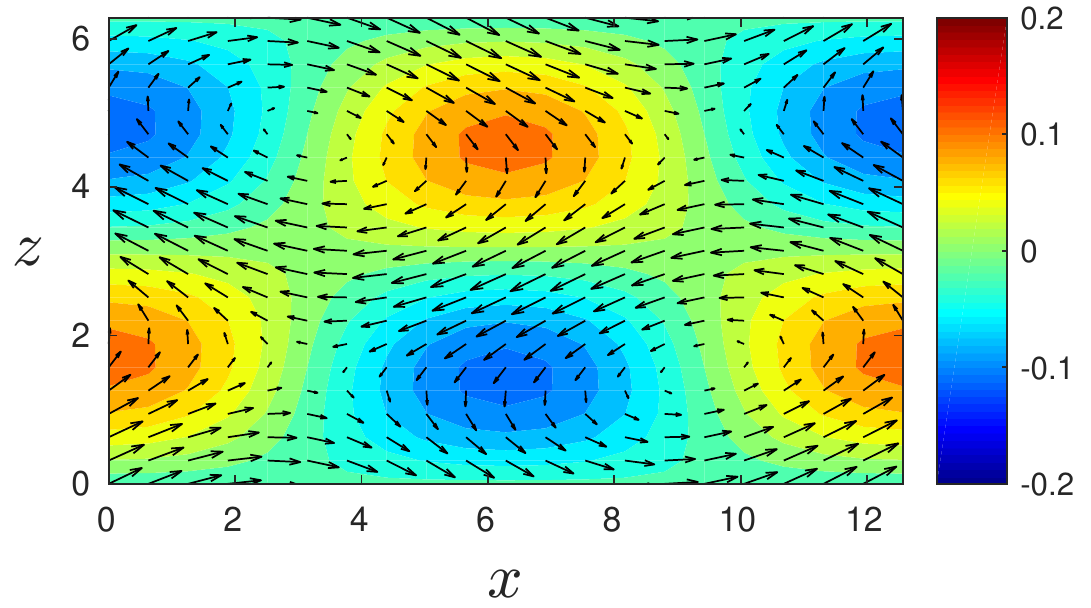}
\caption{Instantaneous velocity fields reconstructed from the mode amplitudes for the LSTM1 architecture. The colored contours represent the velocity component perpendicular to the visualized plane, and the vectors the velocity components in the plane. The left panel is averaged in the streamwise direction, and the right panel shows the flow in the vertical midplane.}
\label{fig_field}
\end{figure}

One of the most relevant features of the nine-equation model by Moehlis {\it et al.}\cite{moehlis_et_al} is the fact that it contains the features present in the near-wall cycle of wall-bounded turbulence,\cite{hamilton_et_al} namely the streamwise vortices, the streaks, their instabilities and the nonlinear interactions among the various structures. Instantaneous velocity fields obtained from the LSTM1 architecture are shown in Figure~\ref{fig_field}, where it can be observed that the network exhibits the streamwise vortices convecting near-wall fluid towards the opposing wall, forming high- and low-speed streaks when the flow is lifted up and pulled down towards the wall, respectively. These streaks become unstable due to a sinusoidal modulation in the spanwise direction, which is visible on the vertical midplane. The close resemblance between the LSTM flow fields and the ones reported by Moehlis {\it et al.}\cite{moehlis_et_al} indicates that the LSTM network reproduces the underlying physics in the nine-equation model, a fact that further justifies the excellent agreement of the turbulence statistics.

\begin{figure}[t!]
\centering
\includegraphics[width=0.95\textwidth]{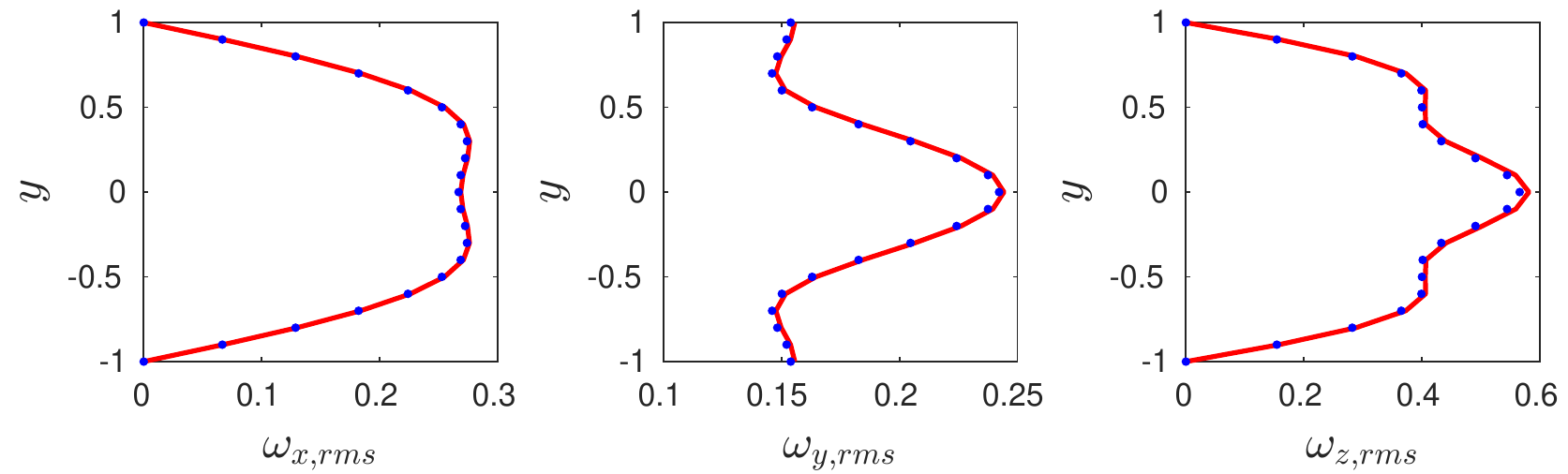}
\caption{\textcolor{black}{(Left) Streamwise, (middle) wall-normal and (right) spanwise root-mean-square vorticity fluctuations obtained from (blue dots) the reference nine-equation model and (red lines) the LSTM1 network. Note that the network was trained with 10,000 datasets.}}
\label{fig_vort}
\end{figure}

\textcolor{black}{An additional assessment of the flow behavior can be obtained in terms of the fluctuating vorticity components. The three components of the instantaneous vorticity were computed by calculating the spatial derivatives of the Fourier modes $\mathbf{u}_{j}(\mathbf{x})$, and using the corresponding time coefficients $a_{j}(t)$. In Figure~\ref{fig_vort} we show the root-mean-square (r.m.s.) vorticity fluctuations in the three spatial directions, obtained from the model and the LSTM1 network trained with 10,000 datasets. It is interesting to note that the nine-equation model exhibits a number of characteristic features from wall-bounded turbulence, such as the location of the relative maxima in the streamwise vorticity fluctuation profile $\omega_{x,rms}$, approximately coinciding with the center of the near-wall streamwise vortices (see Figure~\ref{fig_field}),~\cite{kim_et_al} {\it i.e.} $y \simeq 0.25$ from the centerline. Regarding the LSTM predictions, they are in excellent agreement in the three cases, as in the other turbulence statistics reported in this study, with relative errors of $0.6\%$, $0.7\%$ and $1.3\%$ for the streamwise, wall-normal and spanwise fluctuating vorticities, respectively.}

\begin{figure}[t!]
\centering
\includegraphics[width=1\textwidth]{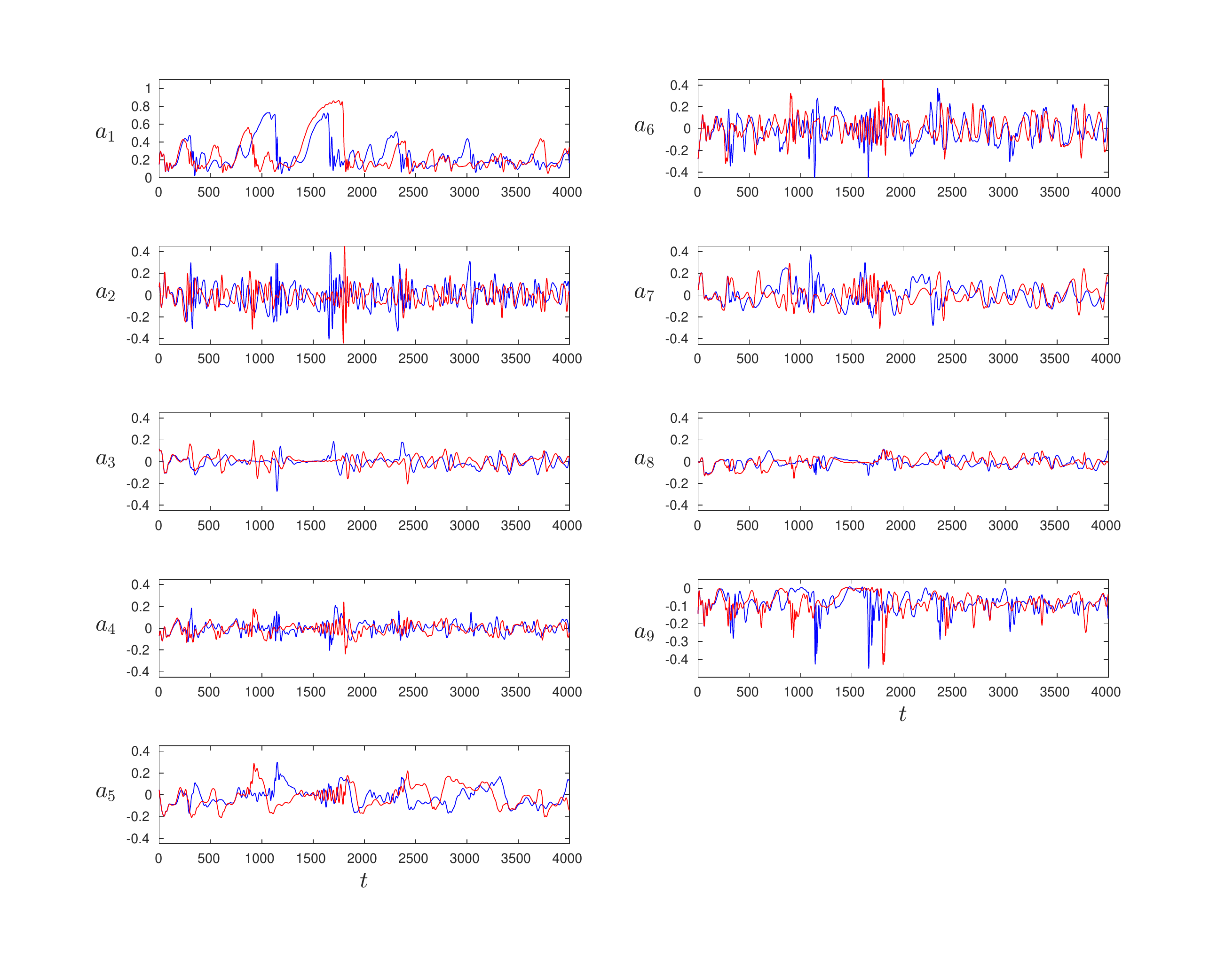}
\caption{\textcolor{black}{Temporal evolution of the nine coefficients for a particular time series, from (blue) the nine-equation model and (red) the LSTM1 network trained with \textcolor{black}{ 10,000} datasets.}}
\label{fig_amp}
\end{figure}

It is important to note that although the LSTM1 network yields excellent predictions of the turbulence statistics and the predicted fields exhibit the most relevant flow structures, it shows deviations in the instantaneous predictions of the mode amplitudes. \textcolor{black}{This can be observed in Figure~\ref{fig_amp}, where the amplitudes of the nine coefficients are predicted for a single time series using the LSTM1 network (trained with 10,000 datasets) and are compared with the reference data from the nine-equation model. As shown in Table~\ref{table_lstm}, we consider a value $p=10$ for this network, which implies that 10 points are used for the prediction. This is evident in the figure, where the 10 first points from each coefficient are identical in the LSTM1 and reference curves. Beyond the first 10 data points, small differences start to emerge in the predicted time series, leading to progressively larger errors and thus different temporal evolutions in all the coefficients. Note however that despite the instantaneous differences in the values of the coefficients, the qualitative behavior of the reference and predicted curves is similar. In fact, certain features appear to be reproduced by the LSTM1 network, such as for instance the large peaks in $a_{1}$ observed at $t\simeq 1,000$ and $1,600$, which are observed with small shifts in time and different amplitudes in the predicted series. Similar observations can be made for the rest of coefficients regarding the most prominent features of the various time series.}

\begin{figure}[t!]
\centering
\vspace{0pt}
\includegraphics[width=0.34 \textwidth]{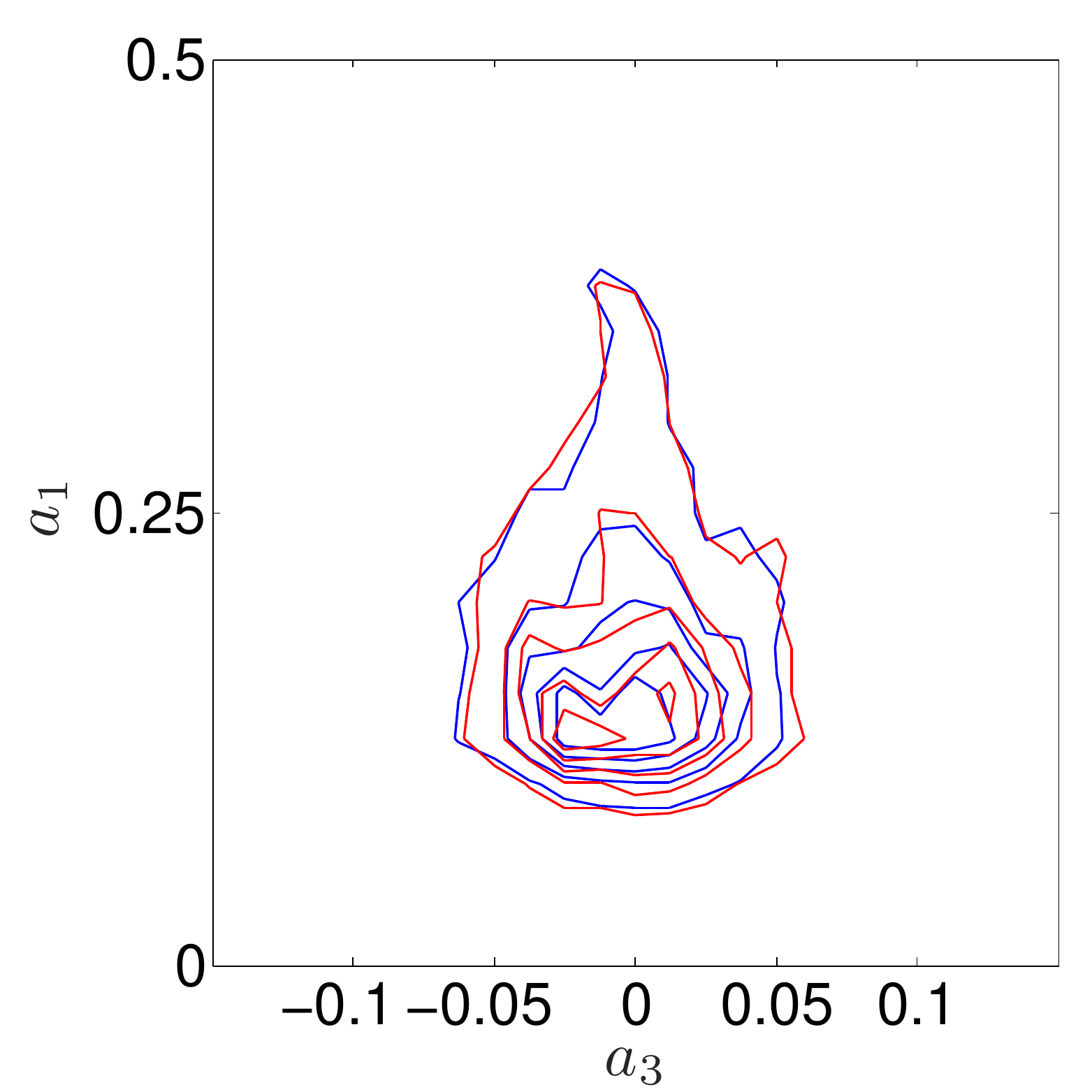}
\includegraphics[width=0.47 \textwidth]{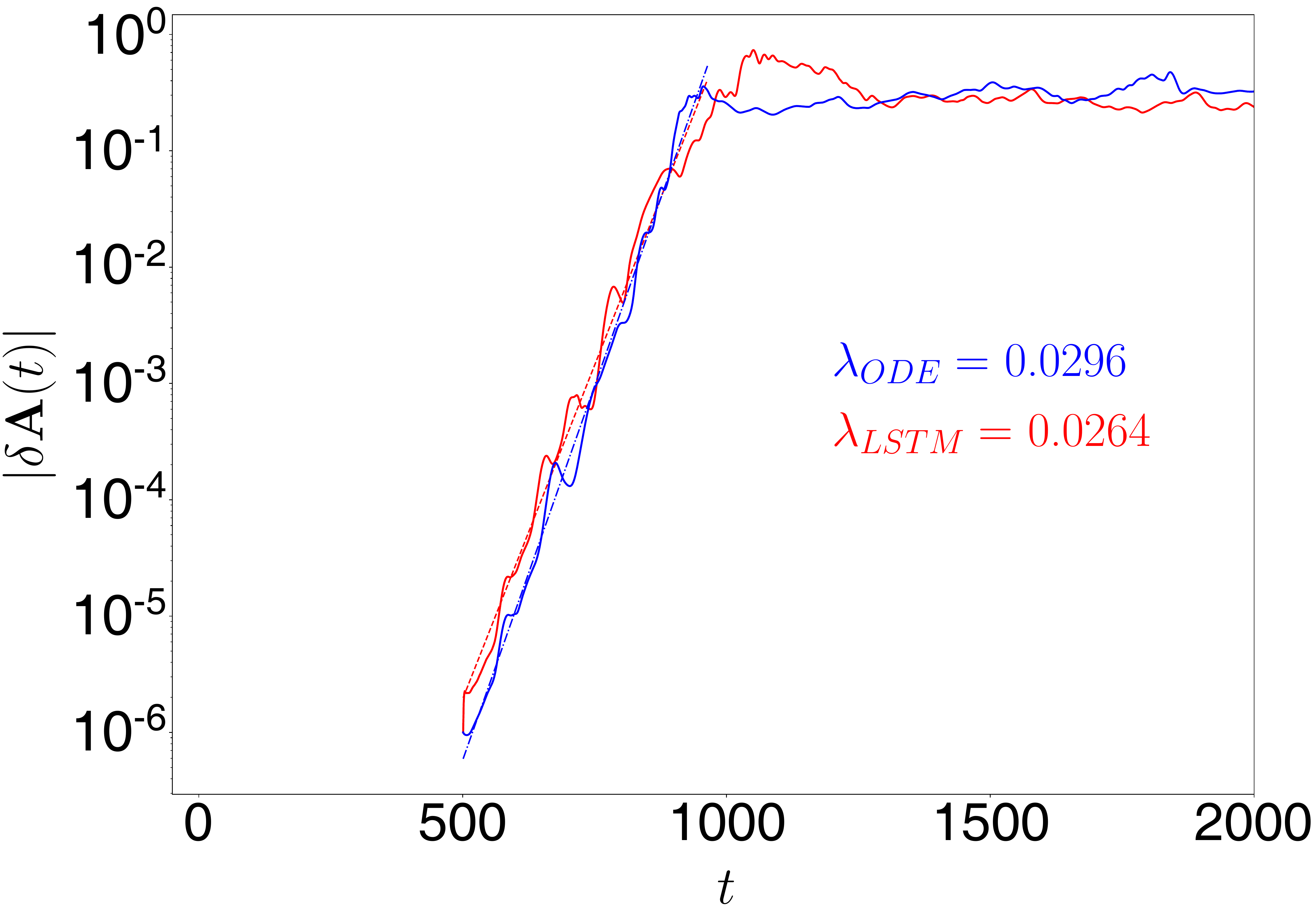}
\caption{(Left)  Probability density function of the Poincar\'e maps, where the intersection with the $a_{2}=0$ plane (with ${\rm d}a_{2}/{\rm d}t <0$) is shown. (Right) Ensemble-averaged divergence of instantaneous time series after a perturbation with $| \delta \mathbf{A}_{0} |=10^{-6}$ is introduced at $t_{0}=500$, showing initial exponential growth and the value of the Lyapunov exponent (dashed lines added to illustrate the obtained slope). In both panels blue and red denote reference model and LSTM prediction, respectively.}
\label{fig_dynamical}
\end{figure}

\textcolor{black}{In order to quantify the deviation between the reference and predicted series discussed above, we} define the relative error in the prediction of the amplitude $a_{1}(t)$ in a single time series consisting of $N_{s}$ samples as: 
\begin{equation}\label{e_eq}
\textcolor{black}{e_{1}=\frac{1}{ (N_{s}-p) a_{1,{\rm lam}} } \sum_{j=p+1}^{N_{s}} \left | a_{1,{\rm mod}}^{j}-a_{1,{\rm pred}}^{j} \right |,}    
\end{equation} 
 where also here `mod' and `pred' denote data from the nine-equation model and predictions from the NN, respectively, and $a_{1,{\rm lam}}=1$ is the value for laminar flow. \textcolor{black}{Note that the difference between equations (\ref{epsilon_eq}) and (\ref{e_eq}) lies in the data used as a reference: whereas in the former we use the training data ({\it i.e.} data seen by the network during training), in the later we use new data which has not been employed for training the network.} Despite the very low validation loss obtained when considering the LSTM1 network trained with 10,000 datasets ($5.2 \times 10^{-9}$), the relative error in the instantaneous prediction of $a_{1}$ was relatively high ($e_{1}=13.08\%$), and those of the other modes were of comparable magnitude. The reason for this discrepancy lies in the nature of the nine-equation model, which is a dynamical system in which very small perturbations lead to completely different temporal evolutions of the modes. Thus, even if the network is able to reproduce the behavior of the flow very accurately, the instantaneous values differ due to the chaotic nature of the system. In order to further analyze the quality of the predictions, we assessed whether the dynamic behavior of the predicted flow was consistent with the reference model. To this end, we first consider the Poincar\'e map defined as the intersection of the flow state with the hyperplane $a_{2}=0$ on the $a_{1}-a_{3}$ space (with the additional condition ${\rm d}a_{2}/{\rm d}t <0$). This Poincar\'e map is a lower-dimensional representation of the system which allows to assess whether the correlation between amplitudes $a_{1}$ and $a_{3}$ (which are the amplitudes of the modes containing the base laminar profile and the streamwise vortices\cite{moehlis_et_al}) is adequately reproduced by the LSTM network. In Figure~\ref{fig_dynamical}~(left) we show the probability density function (pdf) of the Poincar\'e maps constructed from the 500 time series obtained from the LSTM1 prediction and the reference nine-equation model. The similarity between both maps shows that the LSTM network effectively captures the correlation between both amplitudes, suggesting that the interaction between modes $\mathbf{u}_{1}$ and $\mathbf{u}_{3}$ is adequately represented by the network. Additionally, we analyzed the rate of separation among trajectories in the nine-equation model and the LSTM predictions through their Lyapunov exponents. Given two time series 1 and 2, we define the separation of these trajectories as the Euclidean norm in nine-dimensional space: 
 \begin{equation}
\textcolor{black}{\left | \delta \mathbf{A}(t) \right | = \left [ \sum_{i=1}^{9} \left (a_{i,1}(t)-a_{i,2}(t)  \right )^{2} \right ]^{1/2},}
\end{equation} 
and denote the separation at $t=t_{0}$ as $ | \delta \mathbf{A}_{0} |$. The initial divergence of both trajectories can be assumed as:
\begin{equation}
\textcolor{black}{\left | \delta \mathbf{A}(t') \right | = \exp (\lambda t') \left | \delta \mathbf{A}_{0} \right |,}    
\end{equation}
where $\lambda$ is the so-called Lyapunov exponent and $t'=t-t_{0}$. Here we introduce a perturbation with norm $| \delta \mathbf{A}_{0} | =10^{-6}$ (which corresponds to the order of magnitude of the accuracy of the LSTM1 architecture) at $t_{0}=500$, where all the coefficients are perturbed, and analyze its divergence with respect to the unperturbed trajectory. In Figure~\ref{fig_dynamical}~(right) we show the evolution of $ | \delta \mathbf{A}(t) |$ with time for both the nine-equation model and the LSTM prediction, ensemble-averaged over 10 initial conditions. The rates of divergence from both reference and prediction are very similar, and the estimated values of $\lambda$ are almost identical: 0.0296 for the reference and 0.0264 for the LSTM. After the initial period of divergence, both the nine-equation model and the network saturate at $t \simeq 1,000$. This again supports the claim that the LSTM prediction reproduces the dynamical behavior of the original nine-equation model, and the discrepancy in instantaneous predictions is due to the chaotic nature of the system.

\textcolor{black}{\section{Concluding remarks} \label{sec:conclusions}} 

In this study we have illustrated the potential of neural networks to predict the temporal dynamics of \textcolor{black}{ a low-order model of a turbulent flow}. We used the nine-equation model by Moehlis {\it et al.}\cite{moehlis_et_al} to train a number of MLP and LSTM networks, using several architectures, and obtained flow predictions from all of them. While both types of neural network are able to capture the most relevant flow structures present in the flow, the LSTM yields excellent agreement with the reference data, both in terms of turbulence statistics and dynamical behavior of the system. Note that we employed the Moehlis {\it et al.}\cite{moehlis_et_al} model of the near-wall cycle due to its simplicity and flexibility to produce large amounts of training data, thus facilitating the training and testing of the various network architectures. After this initial assessment of the suitability of using neural networks to predict simplified turbulent flows, the next step will be to extend the present machine-learning framework to cases where the governing equations are \emph{not} known, \emph{e.g.}\ the LES equations arising due to non-trivial (spatial and temporal) filtering, or the generation of (instantaneous) inflow and boundary conditions. 

\textcolor{black}{ The present study is aimed at assessing the feasibility of neural network predictions on a low-order representation of near-wall turbulence. Therefore its scope is limited to the model by Moehlis {\it et al.},\cite{moehlis_et_al} where the basis functions were fixed \emph{a priori}, and the network prediction pertained only to the time coefficients. Despite this limitation, the considered system can be viewed as a representative model for the near-wall cycle of wall-bounded turbulence assuming that the flow is properly scaled in inner units for which the near-wall cycle (disregarding outer-layer influences) is universal. On the other hand, the present methods could be extended to more challenging flows such as channels and boundary layers, including advancing from a low-order model to a higher-order representation of the flow dynamics. This step will necessitate the definition of proper basis functions in addition to the temporal dynamics, which could be achieved based on \textit{e.g.}\  Fourier or POD (proper orthogonal decomposition) methods.} 

Using a standard workstation (Intel(R) Core(TM) i7-4930K CPU at 3.4 GHz), the LSTM1 training with 10,000 time series required around 70 hours. After training the network, it takes around 12 minutes to produce 500 time series, which is the amount of data required to obtain converged statistics. On the other hand, on the same workstation it takes around 6 minutes to produce the same amount of data by integrating the nine-equation model by Moehlis {\it et al.}\cite{moehlis_et_al} Thus, after the initial investment required to train the network, the computer time necessary to predict the flow is only around twice as large as that of resolving a nine-equation model. Furthermore, neural networks, once trained, can be accurately summarized into smaller and more computationally efficient networks which will reduce the time complexity of prediction significantly.\cite{hubara2017quantized} Consequently, we consider that the computational cost of evaluating the neural network is sufficiently low to constitute an efficient alternative for predicting instantaneous variables as \emph{e.g.}\ in SGS models. Since the network designs discussed here can also be employed for the prediction of other types of flows, this study may serve as a guideline for the most suitable strategies to perform such predictions. Other natural extensions of the present work include the use of deep neural networks to generate inflow conditions for spatially developing flows (see the study by Jarrin {\it et al.}\cite{jarrin_et_al}), and also to set off-wall boundary conditions in high-Reynolds-number wall-bounded turbulence simulations (as discussed in the work by Mizuno and Jim\'enez\cite{mizuno_jimenez}). Due to the fact that a significant fraction of the computational cost is employed to resolve the near-wall region, the use of adequate off-wall boundary conditions may allow to achieve very high Reynolds numbers in turbulence simulations. These two applications highlight the \textcolor{black}{potential relevance of the methods described in this study.}

\textcolor{black}{\section*{Acknowledgments}}
 Note that all the TensorFlow setups employed in this work are available online.\cite{codes} The authors thank Martin Lellep (University of Marburg) for helpful comments on the original manuscript. We also acknowledge the funding provided by the Swedish e-Science Research Centre (SeRC) and the Knut and Alice Wallenberg (KAW) Foundation. Part of the analysis was performed on resources provided by the Swedish National Infrastructure for Computing (SNIC) at PDC and HPC2N.

\bibliography{ml_bib}

\end{document}

%% file: symbollines.tex
\font\smallfont=cmsy10 at 10truept
\mathchardef\bigCircle="280D

\font\bigfont=cmsy10 at 14.4truept
\mathchardef\tiMes="2902        %

\font\Bigfont=cmsy10 at 17.28truept
\mathchardef\DiaMond="2A05        %
\mathchardef\cirCle="2A0E
\mathchardef\BigCircle="2A0D

\font\Bbigfont=cmsy10 at 24.88truept
\mathchardef\buLLet="2B0F

\def\bigCirc{\raise 0.3ex\hbox{$\bigCircle$}\nobreak$\,$}

\def\Bullet{\raise-0.35ex\hbox{$\buLLet$}\nobreak$\,$}

\def\triangledown{\raise 0.2em\hbox{$\bigtriangledown$}\nobreak$\,$}

\def\minisquare{\hbox{${\vcenter{
               \hrule height 0.3pt \kern-0.4pt
               \hbox{\vrule width  0.3pt height 3.0pt \kern 2.6pt
               \vrule width  0.3pt height 3.0pt} \kern-0.4pt
               \hrule height 0.3pt}}$}}
\def\ssquare{\raise 0.175ex\hbox{${\vcenter{
               \hrule height 0.5truept       \kern-0.25truept
               \hbox{\vrule width 0.5truept height 3.0truept \kern 2.75truept
                     \vrule width 0.5truept height 3.0truept} \kern-0.25truept
               \hrule height 0.5truept}}$}\nobreak$\,$}
\def\squarex{\raise 0.175ex\hbox{${\vcenter{
               \hrule height 0.8truept       \kern-1.80truept
          \hbox{\vrule width 0.8truept height 8.0truept \kern-1.95truept
                \raise 0.8truept\hbox{$\tiMes$}     \kern-6.70truept
                \vrule width 0.8truept height 8.0truept} \kern-0.80truept
               \hrule height 0.8truept}}$}\nobreak$\,$}

\def\sqbull{\raise0.175ex\hbox{\vrule height 1.4ex width 1.6ex depth 0.2ex}\nobreak$\,$}
\def\smsqbull{\raise0.175ex\hbox{\vrule height 0.8ex width 0.9ex depth 0.2ex}\nobreak$\,$}

\def\Diamondplus{${\vcenter{\vcenter{\DiaMond} \kern-10truept
                            \hbox{\vrule width .4truept}\kern -3truept
                            \hrule height .4truept}}$\nobreak$\,$}
\newcount\ndots

\def\drawline#1#2{\raise 2.5truept\vbox{\hrule width #1truept height #2truept}}
\def\moonspace#1{\hskip #1truept}

\def\Dashy{\drawline{4.00}{1.00}}     
\def\dashy{\drawline{4.00}{0.75}}     
\def\thindashy{\drawline{4.00}{0.25}}     
\def\dashyspace{\dashy\moonspace{2}}
\def\Dashyspace{\Dashy\moonspace{2}}
\def\thindashyspace{\thindashy\moonspace{2}}
\def\longdashy{\drawline{8.00}{0.75}} 
\def\thinlongdashy{\drawline{8.00}{0.25}} 
\def\longdashyspace{\longdashy\moonspace{2}}
\def\thinlongdashyspace{\thinlongdashy\moonspace{2}}

\def\dashbox{\hbox{\dashyspace}}  
\def\Dashbox{\hbox{\Dashyspace}}  
\def\dashed{\hbox {\ndots=0 \loop\ifnum\ndots<3\advance\ndots by 1
        \dashbox\repeat\dashy}\nobreak$\,$}       
\def\Dashed{\hbox {\ndots=0 \loop\ifnum\ndots<3\advance\ndots by 1
        \Dashbox\repeat\Dashy}\nobreak$\,$}       
\def\thindashbox{\hbox{\thindashyspace}}  
\def\thindashed{\hbox {\ndots=0 \loop\ifnum\ndots<3\advance\ndots by 1
        \thindashbox\repeat\thindashy}\nobreak$\,$}       
\def\thindash{\hbox {\ndots=0 \loop\ifnum\ndots<3\advance\ndots by 1
        \thindashbox\repeat\thindashy}\nobreak$\,$}       

\def\longdashbox{\hbox{\longdashyspace}}  
\def\thinlongdashbox{\hbox{\thinlongdashyspace}}  
\def\longdash{\hbox {\ndots=0 \loop\ifnum\ndots<3\advance\ndots by 1
        \longdashbox\repeat\longdashy}\nobreak$\,$}       
\def\thinlongdash{\hbox {\ndots=0 \loop\ifnum\ndots<3\advance\ndots by 1
        \thinlongdashbox\repeat\thinlongdashy}\nobreak$\,$}       

